# Optimization theory and application of nano-microscopic properties of dielectric microspheres

Guorong Sui, Fan Liu, Yuehua Zhang, Xiliang Yang, Xiangmei Dong

University of Shanghai for Science and Technology, Shanghai, China

suigr@usst.edu.cn

Abstract: The dielectric microsphere can be directly embedded into the traditional microscope, which can significantly improve the resolution of the microscope and provide a simple and feasible way to break through the diffraction limit of optical imaging system. However, due to the lack of effective theory and formula system, the resolution limit, magnification and optimal imaging position of the microsphere microsystem cannot be determined, which hinders its application. In this paper, the microscopic theory of dielectric microspheres is studied systematically, and the ray optics is extended to the imaging of dielectric microspheres, so as to establish a formula system containing important optical parameters such as resolution, magnification and imaging position, which provides a solid theoretical basis for further optimization of the nano-microscopic properties of dielectric microspheres. Compared with simulation and experiment, the correctness of the formula system is ensured. The formula shows that the mismatch between refractive index and ambient refractive index limits the resolution of microspheres, and it is an effective method to further improve the resolution to find the optimal refractive index ratio. In addition, the larger refractive index of the medium microsphere and the refractive index of the environment are conducive to the enhancement of the imaging magnification, but the working distance of the microscope objective must be taken into account in the experiment to obtain the ideal imaging position. The theoretical system proposed in this paper effectively explains the optical principle of dielectric microsphere imaging, gives optimized optical parameters, and has been applied in experiments, which is of great value for the practical application of dielectric microsphere nanometer microscopy.

## 1 Introduction

Breaking through optical diffraction limit and improving imaging resolution have always been the focus of optical research. With the continuous development of optical instruments and imaging methodology, the resolution of various microscopic imaging has been significantly improved. At present, the resolution of optical near-field scanning microscope [1-3] is higher than 100nm, X-ray microscope [4-5] can resolve spatial information of less than 10nm, atomic force microscopic imaging [6] can reach resolution of 1nm, scanning tunnel-microscope [7] can resolve detailed structure of 0.1nm, and electron transmission microscope [8] can reach resolution of pico-meter. However, all kinds of microimaging above have some disadvantages. Optical near-field scanning microscopes, atomic force microscopes, and scanning tunneling microscopes all require scanning probes that can easily damage samples. The photon energy of X-ray microscope light source is too large, and it is also easy to damage the sample. Scanning tunneling microscope (STM) and electron transmission microscope (tem) both require vacuum operation and are environmentally demanding.

Moreover, none of these microscopic images allow real-time observation of living samples. In 2011, wang [9] observed attached silica microsphere samples with a traditional microscope and obtained images with a resolution of 50nm, which opened the nanometer microscopic research on dielectric microspheres. Because the method is simple to operate and has strong real-time performance, it provides a new way for real-time imaging of living samples, so it has attracted extensive attention in the field of optical super-resolution microscopy.

Recently, Li[10] combined medium microsphere with laser scanning confocal microscope to obtain imaging resolution of 25nm.In order to further improve the resolution of nano-microscope of medium microspheres, the influence of geometric and physical parameters of microspheres and the special structure of microsphere evolution on the imaging resolution has been extensively studied [11-14], and Hao [15], Lee[16] and Arash Darafsheh [17-18] et al. found that the immersion degree of immersion solution also affects the imaging resolution of microspheres. In addition, some researches focus on the mechanism of microsphere microimaging, hoping to fundamentally break the limitation of imaging resolution. Existing media microsphere microimaging theories mainly include rice scattering theory [19-20], pattern coupling theory [21], nano energy flow theory [22-24] and evanescent field theory [25-27].All these theories treat the imaging problem of medium microsphere from the perspective of the propagation of light wave electromagnetic field, apply electromagnetic mode decomposition and specific boundary conditions to solve the electromagnetic field distribution at the imaging end, and then obtain the relevant characteristic parameters of the image. However, due to the lack of deterministic description of object-image relationship, important optical performance parameters including imaging resolution, magnification and imaging position cannot be theoretically optimized. Moreover, the quantitative study on the influence of refractive index on the microscopic properties of medium microspheres and the limit of microsphere microscopic imaging resolution are not enough. Therefore, this article established the dielectric microsphere microscopic imaging theory, derived the formulas of imaging performance parameters system, and detailed research on dielectric microspheres nano microscopic imaging performance, especially the refractive index with medium microspheres and environment refractive index matching method to optimize the imaging resolution, magnification and the imaging position, and obtained the experimental verification.

In this paper, the microscopic theory of dielectric microspheres is studied systematically, and the ray optics is extended to the imaging of dielectric microspheres, so as to establish a formula system containing important optical parameters such as resolution, magnification and imaging position, which provides a solid theoretical basis for further optimization of the nano-microscopic properties of dielectric microspheres. Compared with simulation and experiment, the correctness of the formula system is ensured. The formula shows that the mismatch between refractive index and ambient refractive index limits the resolution of microspheres, and it is an effective method to further improve the resolution to find the optimal refractive index ratio. In addition, the larger refractive index of the medium microsphere and the refractive index of the environment are conducive to the enhancement of the imaging magnification, but the working distance of the microscope objective must be taken into account in the experiment to obtain the ideal imaging position. The theoretical system proposed in this paper effectively explains the optical principle of dielectric microsphere imaging, gives optimized optical parameters, and has been applied in experiments, which is of great value for the practical application of dielectric microsphere nanometer microscopy.

**2 Principle of microsphere microimaging**

The dielectric microsphere covering the surface of the sample is like a hyperlens, which can effectively converge the light of the object, increase the focal degree of the microscopic system, and thus improve the spatial resolution of the microscopic imaging. Previous studies have shown that in the imaging field covered by medium microsphere, the imaging resolution can be obtained which is better than that of traditional microscope, and the imaging characteristics meet the basic geometric relationship [9]. In order to further clarify the optical principle of microsphere microimaging and the relationship between object and image, the spherical refraction surface model is used for detailed analysis and discussion. As shown in figure 1, the curvature radius of the front and back refraction surfaces of $SiO_2$ sphere is $r_1=R (r_1>0)$ and $r_2=-r (r_2<0)$, respectively, the thickness is $D=2R$, the refractive index is $n_d$, and the surrounding refractive index is $n_{s1}$ and $n_{s2}$, respectively.

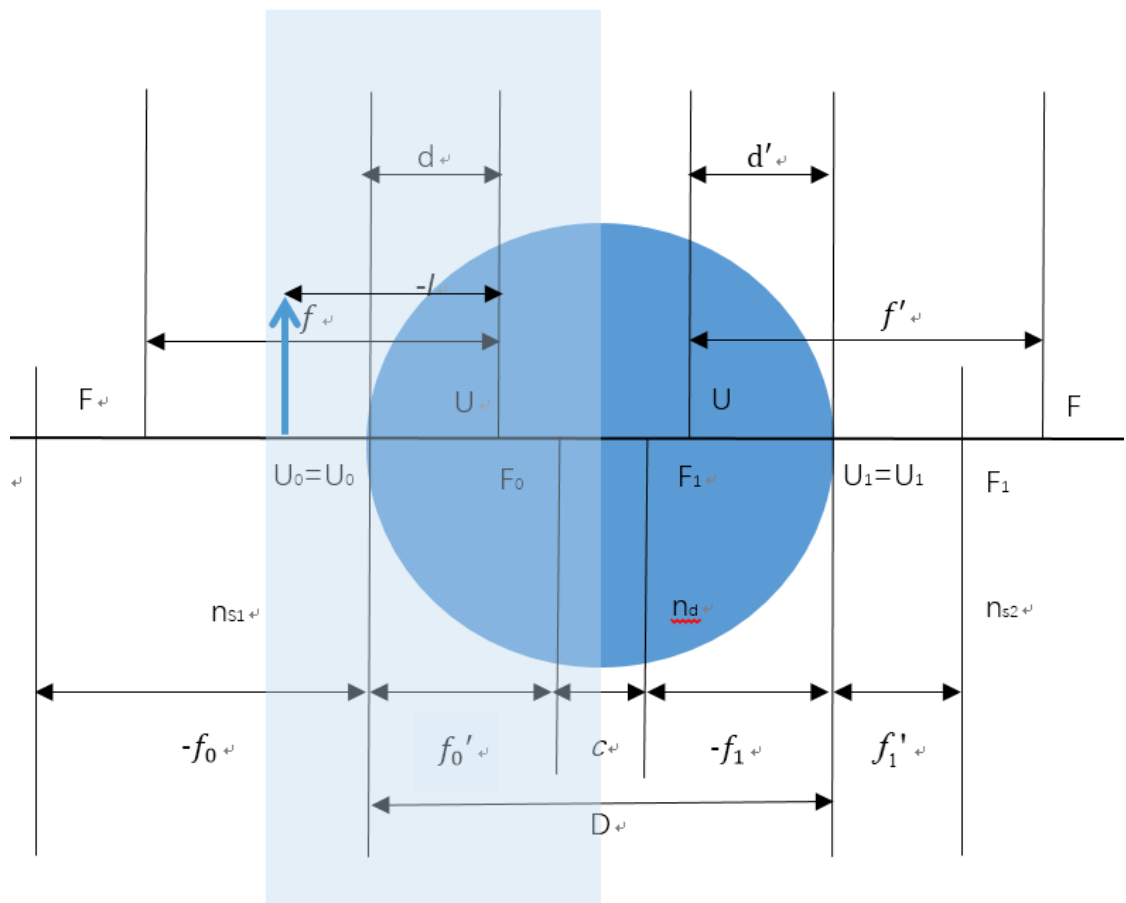

Figure 1. imaging schematic diagram of medium microsphere

According to the ray optics principle of the dielectric interface, the front and back focal lengths of the spherical front surface are respectively:

$$f_0 = -\frac{n_{s1} r_1}{n_d - n_{s1}}, \quad f_0' = \frac{n_d r_1}{n_d - n_{s1}} \quad (1)$$

The front focal length and back focal length of the sphere's rear surface are respectively:

$$f_1 = -\frac{n_d r_2}{n_{s2} - n_d}, \quad f_1' = \frac{n_{s2} r_2}{n_{s2} - n_d} \quad (2)$$

The combined front and back focal lengths of the sphere are respectively:

$$f = -\frac{f_0 f_1}{c}, \quad f' = \frac{f_0' f_1'}{c} \quad (3)$$

Where, c is the distance between the front focus F0 'of the front surface and the front focus F1 of the back surface, which can be expressed as:

$$c = D - f_0' + f_1 \quad (4)$$

Where, D is the axial thickness of the lens, namely the distance between U0 and U1. Substitute equation (1) and equation (2) into equation (4) and get:

$$c = \frac{T}{(n_d - n_{s1})(n_{s2} - n_d)} \quad (5)$$

here:

$$T = -(n_d - n_{s1})(n_{s2} - n_d)d - n_d[(n_{s2} - n_d)r_1 + (n_d - n_{s1})r_2] \quad (6)$$

Substitute equations (1), (2) and (5) into equation (3) and get:

$$f = n_{s1} n_d \frac{r_1 r_2}{T}, \quad f' = -n_d n_{s2} \frac{r_1 r_2}{T} \quad (7)$$

The distance from the pole U0 to the main point U and the distance from the pole U1 to the main point U 'are respectively:

$$d = n_{s1}(n_{s2} - n_d)\frac{r_1 D}{T}, \quad d' = -n_{s2}(n_d - n_{s1})\frac{r_2 D}{T} \quad (8)$$

In practice, the immersion fluid around the microsphere usually has two types: total immersion and semi-immersion. In a full-immersion system, the index of refraction of the immersed liquid ns=ns1=ns2, the radius of curvature of the front surface r1=R, and the radius of curvature of the rear surface r2= -r. Substitute these parameters into equations (6), (7), and (9):

$$T = 2R(n_s - n_d)n_s \quad (9)$$

$$f' = \frac{n_d R}{2(n_d - n_s)} = -f \quad (10)$$

$$d = R, d' = -R \qquad (11)$$

When both object square light and image square light are immersed in liquid, the object image distance relation can be written as:

$$\frac{1}{l'} - \frac{1}{l} = \frac{1}{f'} \qquad (12)$$

Axial magnification is:

$$\beta = \frac{l'}{l} \qquad (13)$$

In general, the medium microsphere is close to the sample surface, so l= -r is substituted into equation (12), and the phase position is obtained as:

$$l' = \frac{Rn_d}{n_d - 2n_s} \qquad (14)$$

Axial magnification is:

$$\beta = \frac{n_d n_s}{2n_s^2 - n_d n_s} = \frac{\dfrac{n_d}{n_s}}{2 - \dfrac{n_d}{n_s}} \qquad (15)$$

Numerical aperture (NA) [28] is:

$$NA = \frac{n_s}{\sqrt{1 + \left[\dfrac{n_d}{2(n_d - n_s)}\right]^2}} \qquad (16)$$

The half-height full width of the point diffusion function [29] is:

$$dt = \frac{0.375\lambda}{NA} \qquad (17)$$

Substitute equation (16) into equation (17) to get:

$$dt = \frac{0.375\lambda \sqrt{1 + \left[\dfrac{n_d}{2(n_d - n_s)}\right]^2}}{n_s} \qquad (18)$$

In the semi-immersed system, the index of refraction of the immersed liquid is ns1, the index of refraction of the air is ns2, the radius of curvature of the front surface r1=R, and the radius of curvature of the rear surface r2= -r. Substitute these parameters into equations (6), (7), and (9):

$$T = -R(n_d n_{s2} - 2n_{s1} n_{s2} + n_{s1} n_d) \qquad (19)$$

$$f = \frac{-Rn_{s1} n_d}{n_d n_{s2} - 2n_{s1} n_{s2} + n_{s1} n_d}, \quad f' = \frac{Rn_d n_{s2}}{n_d n_{s2} - 2n_{s1} n s_2 + n_{s1} n_d} \qquad (20)$$

$$d = \frac{2R(n_{s1}n_d - n_{s1}n_{s2})}{n_d n_{s2} - 2n_{s1}n_{s2} + n_{s1}n_d}, \quad d' = -\frac{2R(n_d n_{s2} - n_{s1}n_{s2})}{n_d n_{s2} - 2n_{s1}n_{s2} + n_{s1}n_d} \tag{21}$$

Since the object square ray and the image square ray in the semi-immersed system are in different media, the object image distance formula is:

$$\frac{f'}{l'} + \frac{f}{l} = 1 \tag{22}$$

Axial magnification is:

$$\beta = -\frac{fl'}{f'l} \tag{23}$$

When the distance from microsphere center to sample is $l = \frac{3R}{4}$, then:

$$l = -\frac{2R(n_{s1}n_d - n_{s1}n_{s2})}{n_d n_{s2} - 2n_{s1}n_{s2} + n_{s1}n_d} - \frac{R}{4} \tag{24}$$

Substitute equation (24) into equation (21) to obtain the phase position:

$$l' = -\frac{2R(n_d - n_{s2})}{n_d - 4n_{s1} - 2n_{s2} + \frac{n_{s1}n_d}{n_{s2}} + 4\frac{n_{s1}n_{s2}}{n_d}} \tag{25}$$

Axial magnification is:

$$\beta = -\frac{1}{1 + \frac{-2 - 2\frac{n_{s2}}{n_{s1}} + 4\frac{n_{s2}}{n_d}}{-2 + \frac{n_d}{n_{s1}} + \frac{n_d}{n_{s2}}}} \tag{26}$$

The half-height full width of the point diffusion function is:

$$dt = \frac{0.375\lambda \sqrt{1 + \left[\frac{n_d n_{s2}}{2n_{s1}n_{s2} - n_{s1}n_d - n_d n_{s2}}\right]^2}}{n_{s2}} \tag{27}$$

## 3 Result and simulation

In order to ensure the correctness of the theory, FDTD simulation algorithm and physical experiments were used to make a comparative study on the media microsphere imaging with different parameters. The focal spot size of the ideal point light source after microsphere focusing directly determines the image resolution of the microsphere. Therefore, it is of great significance to study the effect of the refractive index difference between the microsphere and the surrounding environment on the focal spot size. The diameter of the microsphere used in the simulation is 5 microns, the incident light wave is linearly polarized plane wave, and the plane wave source is placed 5.5 microns away from the center of the microsphere. As shown in figure 2b, when nd-ns>0.8, most of the light energy gathered in the inside of the medium microsphere and formed a small focusing spot, only a small part of the energy ejected out of the outside of the sphere. This indicates that a large refractive index difference will cause a strong local resonance effect in the microsphere, which is not conducive to the image observation. When 0<nd-ns<0.85, the focal point is located outside the dielectric microsphere. At this time, the medium microsphere is equivalent to an ultra-micro focal lens, and its imaging performance is very similar to that of an ordinary spherical lens, as shown in figure 2c. When the refractive index of the medium microsphere is greater than the refractive index of the surrounding environment (nd-ns<0), the focusing spot cannot be obtained on the inside and outside of the medium microsphere, and the medium microsphere shows an obvious edge diffraction effect, as shown in figure 2d.

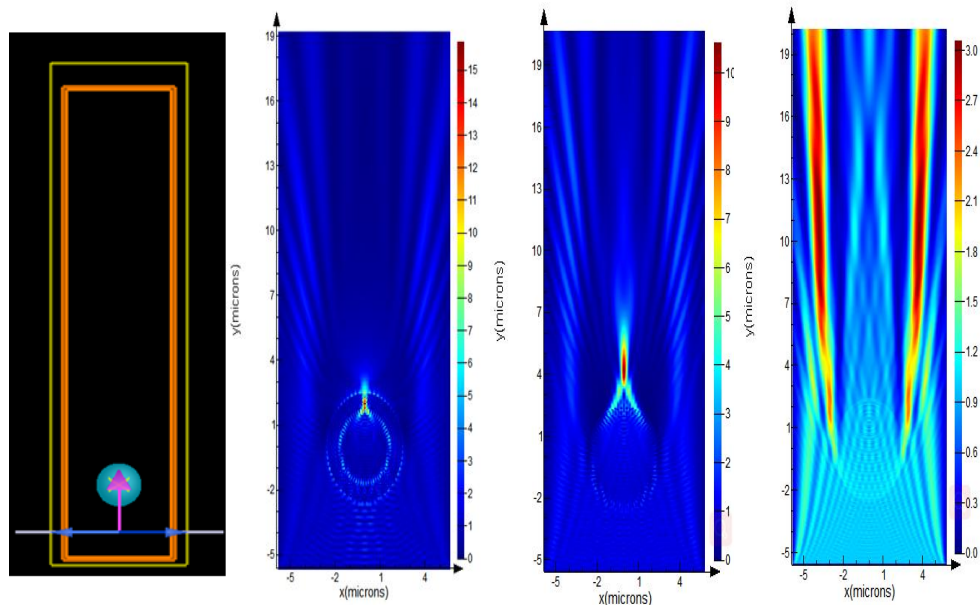

Figure 2. Light field distribution of plane wave lighting medium microsphere (diameter of the simulated medium microsphere is 5 microns, a) simulation structure b)nd-ns > 0.85, c) 0.85>nd-ns>0, d)nd-ns<0)

In order to further study, the imaging resolution of medium microspheres, the following is a detailed analysis of the relationship between the focal spot size and the refractive index of the surrounding microsphere. Take the commonly used barium titanate microsphere as an example, its refractive index is 1.9. As shown in figure 3, the focusing spot size of the medium microsphere changes significantly with the increase of the refractive index of the surrounding environment, and presents a similar concave curve. In addition, the larger the surrounding refractive index is, the larger the focal spot size will be, which shows obvious nonlinearity. When the refractive index of the microsphere is 1.9, the incident light wavelength is 450nm, the semi-peak width of the transverse point diffusion function is 238nm, and the resolution can reach 160nm. When the refractive index of microsphere is determined, the refractive index of immersed liquid has a great influence on the resolution. No matter simulation or theoretical derivation, it can be seen that the refractive index and resolution of the immersed liquid are not monotonically increasing. When the refractive index of the microsphere is 1.9, no matter what range of incident light wavelength is, the refractive index of the immersed liquid is 1.2, which can achieve a relatively small resolution.

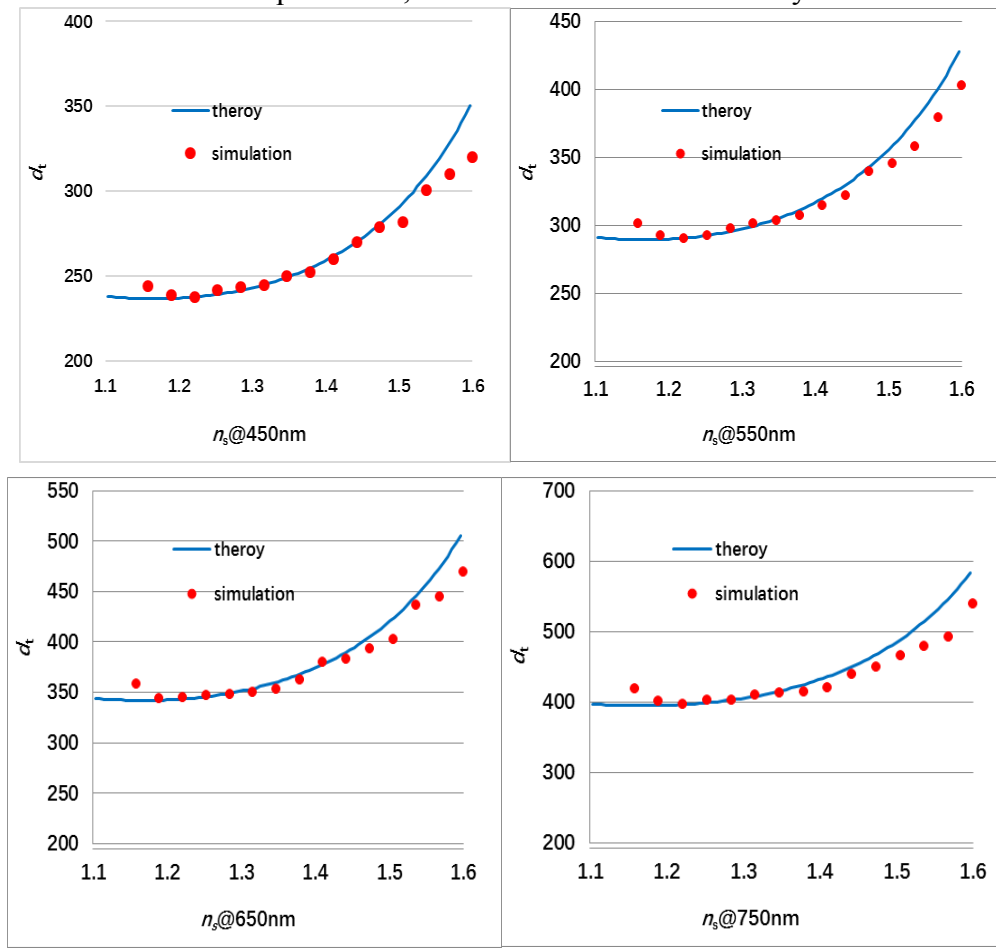

Figure 3. Change of half-height and full-width of point diffusion function under different conditions

When the refractive index of the microsphere is 1.9, the wavelength is: a)450nm, b)550nm, c)650nm, d)750nm, respectively. The blue curve is calculated by the theoretical model formula (18), and the red point is calculated by FDTD simulation:

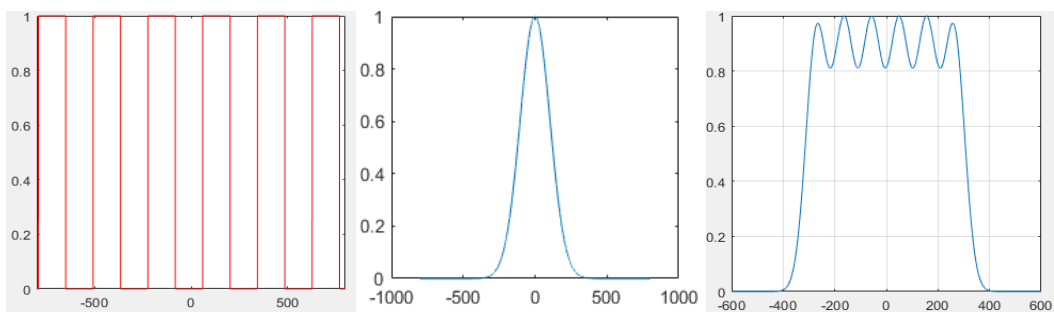

Figure 4. Square wave with slit width of 142nm and peak width of 142nm was convolved with gaussian light with FWHM=238nm.

a) is square wave with period of 284nm, slit width of 142nm and peak width of 142nm, convolved with gaussian function of FWHM=238nm in b), c) is the result after convolution, and the ratio of the minimum and maximum light intensity is 0.81.

According to formula (25) in the theoretical model, it is concluded that the magnification of the experimental image is related to the ratio of nd to ns2 (refractive index ratio = nd/ns2). The larger the ratio is, the larger the magnification of the image is. In optical imaging, magnification is an important index to measure imaging. In this paper, the influence of the relationship between environmental refractive index and microsphere refractive index on imaging magnification is studied theoretically and experimentally. Microscope used in the experiment for Axio Imager. M2 Carl Zeiss microscope (Carl Zeiss) is, numerical aperture NA is 0.9, the emission of light wave as the center wavelength is 550 nm white plane wave. The experiment USES the blu-ray discs (blu-ray disc) as samples, and the cycle of stripe on blu-ray disc is 320 nm, the groove width is 120 nm, grooves between the interval is 200 nm. Place the diameter of 5 um silica microspheres strip on the blu-ray disc of transparent film, adding different refractive index of the immersion fluid, the microsphere was semi-immersed, and the experimental figure measured was shown in Fig. 4.

Table 1. Experimentally Determined Image Magnifications in the Different Media

|  | Super-pure water | Anhydrous ethanol | isopropanol | Immersion oil |
|---|---|---|---|---|
| Refractive index | 1.33 | 1.366 | 1.39 | 1.4 |
| magnification | 2.4 | 2.33 | 2.05 | 1.6 |
| Theoretical magnification | 1.42 | 1.34 | 1.28 | 1.26 |

In the experiment, the refractive index of microspheres is certain, and their refractive index ratio is changed by changing the refractive index of the immersed liquid. The theoretical magnification in table 1 is calculated by formula (25). Through experiments and theories, it can be concluded that the magnification of the experimental image is related to the ratio of nd to ns2 (refractive index ratio = nd/ns2). The larger the ratio is, the larger the magnification of the image is. Our theoretical model allows qualitative analysis of magnification. In the experiment, if the resolution is allowed, the immersion solution with high refractive index can be selected as far as possible.

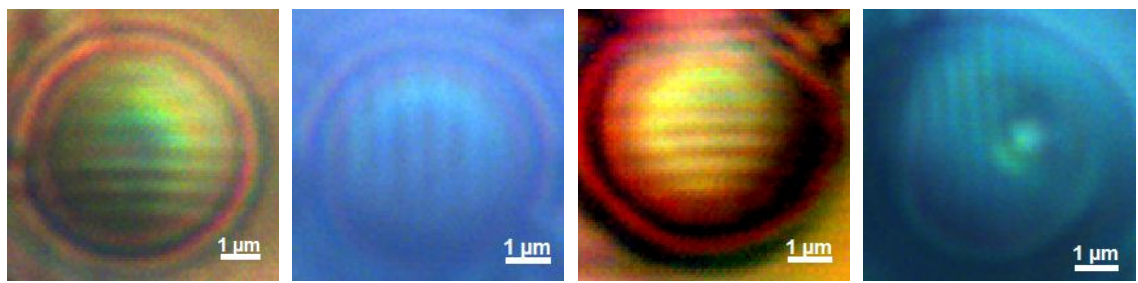

Figure 5. The microspheres are made of silica, and the immersion liquid used is a) anhydrous ethanol b) propanol c) isopropanol d) ionized water, respectively, and the immersion state is semi-immersion

According to the theoretical model, the imaging position of the microsphere can be calculated, and the imaging position can be judged to meet the scope of microscope imaging, which can predict whether the objective lens we use can observe the sample. The microscope we used has a 100-fold objective imaging distance of 310um and a 50-fold objective imaging distance of 570um.Since the microsphere was imitated as a virtual image in the experiment, the image plane was behind the sample. When the upper surface of the microsphere was close to the lens of the objective lens, it might not be possible to make the image plane formed by the microsphere within the object distance of the microscope objective lens. Therefore, it is necessary to pay attention to whether the imaging position of the microsphere meets the object distance of the objective lens. According to our model, the object distance is the relationship between the refractive index ratio and the theoretical data is obtained from formula (14)., when the refractive index ratio is larger, the imaging position is closer to the center of the microsphere and the magnification is smaller. The same experimental conclusion was drawn in Lee's paper, as shown below:

Table 2. Lee's experiment results

|  | water | 40% Sugar Solution | Microscope Immersion Oil |
|---|---|---|---|
| Refractive index | 1.33 | 1.399 | 1.518 |
| Focal image range | 70---150 | 42—87 | 15--45 |
| thorey | 125 | 105.79 | 83.6268 |

**4 Conclusions**

    The model system can be used to predict the optimal resolution of the microsphere imaging system and the ratio of the refractive index of the system with the best resolution, and to provide reference for the magnification of the microsphere and the imaging position of the microsphere. When using barium titanate microspheres with a refractive index of 1.9, the immersion solution with a refractive index of 1.3--1.4 can achieve a better resolution. Under the requirement of resolution, the ratio with a relatively small refractive index between the microsphere and the immersed liquid can achieve a larger magnification. However, it is necessary to calculate whether the imaging position meets the object distance of the microscope objective lens, otherwise the object cannot be imaging. The theoretical system can provide guidance for further study of microsphere imaging experiment.